\documentclass{article}

\RequirePackage{hyperref}
\usepackage{amsmath}
\usepackage{amssymb}
\usepackage{tikz}
\usepackage{listings}
\usepackage[group-separator={,}]{siunitx}
\usepackage{tikz}
\usepackage{tikz-qtree}
\usetikzlibrary{backgrounds,positioning,shapes,arrows,automata,decorations.pathreplacing,angles,quotes}

\newtheorem{defn}{Definition}[section]

\newcommand{\R}{{I\!\!R}}
\newcommand{\V}{{\bf v}}
\newcommand{\X}{{\bf x}}
\newcommand{\Z}{{\bf z}}
\newcommand{\Y}{{\bf y}}

\newcommand{\fma}{{\small {\tt fma}}}

\newtheorem{theorem}{Theorem}

\title{Hessian Chain Bracketing}

\author{Uwe Naumann\thanks{Computer Science, RWTH Aachen University, Germany; naumann@stce.rwth-aachen.de; corresponding author} and
Shubhaditya Burela\thanks{shubhaditya.burela@rwth-aachen.de}}

\begin{document}

\maketitle

\begin{abstract}
	Second derivatives of mathematical models for real-world phenomena are
fundamental ingredients of a wide range of numerical simulation methods
including parameter sensitivity analysis, uncertainty quantification,
nonlinear optimization and model calibration. The evaluation of such
Hessians often dominates the overall computational effort.
The combinatorial {\sc Hessian Accumulation} problem aiming to minimize the
number of floating-point operations
required for the computation of a Hessian turns out to be NP-complete.
We propose a dynamic programming formulation for the solution of
{\sc Hessian Accumulation} over a sub-search space.
This approach yields improvements by factors of ten and higher over 
the state of the art based on
second-order tangent and adjoint algorithmic differentiation.
\end{abstract}

\section{Motivation and Introduction} \label{sec:1}
We consider twice differentiable multivariate vector functions 
$$F : \R^n \rightarrow \R^m : \X \mapsto \Y=F(\X)$$
implemented as computer programs evaluating
sequences of $q>0$ {\em elemental functions} 
$$F_i : \R^{n_{i-1}} \rightarrow \R^{n_i} : \V_{i-1} \mapsto \V_i=F_i(\V_{i-1})$$ 
for $i=1,\ldots,q,$ $\V_0=\X$ and $\Y=\V_q.$ This layered structure of 
\begin{equation} \label{e1.1}
F = F_{q} \circ F_{q-1} \circ F_{q-2} \circ \cdots \circ F_{1} 
\end{equation}
is typical for many numerical simulations. Even if it is not explicit in the
given source program finding suitable vertex separators representing 
the $\V_i$ in the directed acyclic data dependence graph is straightforward.
We set $F_{[i,j)} \equiv F_i \circ \cdots \circ F_{j+1}$ implying
$F_i=F_{[i,i-1)}$ and $F=F_{[q,0)}.$ We use $=$ to denote mathematical equality and $\equiv$ in the sense of ``is defined as.''
Elemental 
Jacobians $$F'_i = F'_i(\V_{i-1}) \equiv \frac{d F_i}{d \V_{i-1}}(\V_{i-1}) \in \R^{n_i \times n_{i-1}}$$ and 
Hessians $$F''_i = F''_i(\V_{i-1}) \equiv \frac{d^2 F_i}{d \V_{i-1}^2}(\V_{i-1}) \in \R^{n_i \times n_{i-1} \times n_{i-1}}$$ are assumed to 
be given. For example, they can be computed by application of
Algorithmic Differentiation (AD) \cite{Griewank2008EDP,Naumann2012TAo}
to a given implementation of the $F_i$ as a differentiable subprogram.

The chain rule of differential calculus yields 
\begin{equation} \label{e1.2}
F' = F'(\X) = F'_q \cdot F'_{q-1}\cdot\ldots\cdot F'_1
\end{equation}
where $$F'_i = F'_i(\V_{i-1}) \equiv \frac{d F'_i}{d \V_{i-1}}(\V_{i-1}) \in \R^{n_i\times n_{i-1}}$$ denotes the Jacobian of $F_i$ for $i=1, \ldots, q$. 
The corresponding Hessians are denoted as $$F''_i = F''_i(\V_{i-1}) \equiv \frac{d^2 F'_i}{d \V_{i-1}^2}(\V_{i-1}) \in \R^{n_i\times n_{i-1} \times n_{i-1}} \; .$$ 
Differentiation of Equation~(\ref{e1.2}) with respect to $\X$ yields 
\begin{equation}\label{e1.3}
\begin{split}
	\left [F'' \right ]&_{\delta,\alpha_1,\alpha_2}
	=\sum_{j=1}^q \left [F'_{[q,j)} \right ]_{\delta,\gamma} \cdot \left [F''_j \right ]_{\gamma,\beta_1,\beta_2} \cdot \left [F'_{[j-1,0)} \right ]_{\beta_1,\alpha_1} \cdot \left [F'_{[j-1,0)} \right ]_{\beta_2,\alpha_2} \\
	&=\sum_{j=1}^q \left  [\prod_{i=j+1}^q F'_i \right ]_{\delta,\gamma} \cdot \left [F''_j \right ]_{\gamma,\beta_1,\beta_2} \cdot \left [\prod_{k=1}^{j-1} F'_k \right ]_{\beta_1,\alpha_1} \cdot \left [\prod_{k=1}^{j-1} F'_k \right ]_{\beta_2,\alpha_2} \; .
\end{split} 
\end{equation}
We use index notation for tensor products. Tensors are enclosed in square 
brackets and summation runs over the common index.
Jacobians and Hessians of subchains of Equation~(\ref{e1.1}) are denoted as
$$
F'_{[i,j)} \equiv \frac{d F_{[i,j)}}{d \V_j} \in \R^{n_i \times n_j}
\quad 
\text{and}
\quad 
F''_{[i,j)} \equiv \frac{d^2 F_{[i,j)}}{d \V^2_j} \in \R^{n_i \times n_j \times n_j} \; .
$$
In the following we use the simplified notation
$$
	F'' 
	=\sum_{j=1}^q \left ( F'_{[q,j)} \cdot F''_j \cdot F'_{[j-1,0)}  \otimes F'_{[j-1,0)} \right ) \; ,
$$
where $\otimes$ denotes the outer product of two matrices as defined in Equation~(\ref{e1.3}).

Different approaches to the evaluation of 
Equation~(\ref{e1.3}) yield varying computational complexities in terms of
the number of scalar {\em fused multiply-add} (\fma) operations required.
The minimization of this cost can be stated formally a combinatorial 
optimization problem yielding the
the following formulation as a decision problem.
\begin{defn}[{\sc Hessian Accumulation}]
Given are a layered twice differentiable function $F$ as 
in Equation~(\ref{e1.1}) together with elemental Jacobians $F'_i$ and 
Hessians $F''_i$ for $i=1,\ldots,q$ and
a positive integer $k\geq 0$. Can the Hessian $F''$ of $F$ be evaluated
with at most $k$ \fma\ operations?
\end{defn}
\begin{theorem} \label{np}
{\sc Hessian Accumulation} is NP-complete.
\end{theorem}
The proof can be found in Section~\ref{a1} of the appendix. It exploits 
potential algebraic dependences among the entries of the elemental Hessians
(equality in particular). The following heuristic assumes these entries to 
be mutually independent (distinct).
We propose a dynamic programming
\cite{Bellman1957DP,Godbole1973}
method for {\sc Hessian Chain Bracketing} formally defined as a 
combinatorial optimization problem as follows:
\begin{defn}[{\sc Hessian Chain Bracketing}]
Given a layered twice differentiable function as 
in Equation~(\ref{e1.1}) together with elemental Jacobians and Hessians,
determine a bracketing of Equation~(\ref{e1.1})
such that the number of \fma\ operations required by Equation~(\ref{e1.3}) 
becomes minimal.
\end{defn}

\paragraph*{Example} To illustrate the potential for optimized instances of
{\sc Hessian Chain Bracketing} consider 
$
F=F_3 \circ F_2 \circ F_1
$
such that $F_3,F_1 \in \R^n \rightarrow \R^m$ and $F_2 \in \R^m \rightarrow \R^n.$
Hence,
$F'_3,F'_1 \in \R^{m \times n},$ $F_2 \in \R^{n \times m}$ and
$F''_3,F''_1 \in \R^{m \times n \times n},$ $F_2 \in \R^{n \times m \times m}.$
Without loss of generality, all elemental Jacobians and Hessians are assumed 
to be dense. Tracking of highly likely sparsity would complicate the 
presentation of the example while not offering any further conceptual insight.

There are two ways to split $F$ yielding the following
$\fma$ costs
\begin{itemize}
	\item $F=F_3 \circ (F_2 \circ F_1):$ From
		$
F''=F''_{[3,0)} = F'_3 \cdot F''_{[2,0)} + F_3'' \cdot F'_{[2,0)} \otimes F'_{[2,0)}$
with
$F'_{[2,0)}= F'_2 \cdot F'_1 $ and $F''_{[2,0)} = F'_2 \cdot F''_1  + F''_2 \cdot F'_1\otimes F'_1$
it follows that
	$\fma(F'_{[2,0)}) = mn^2 $ and $\fma(F''_{[2,0)}) = 2mn^3 +m^2n^2$
	and hence
$
	\fma(F'') =\fma(F''_{[3,0)})= 5mn^3 +m^2n^2+mn^2.$
\item $F=(F_3 \circ F_2) \circ F_1:$ From 
		$
		F''=F''_{[3,0)} = F''_{[3,1)} \cdot F'_1 + F'_{[3,1)} \cdot F'_1 \otimes F'_1$
with
$F'_{[3,1)} = F'_3 \cdot F'_2$ and 
$F''_{[3,1)} = F'_3 \cdot F''_2  + F''_3 \cdot F'_2\otimes F'_2$
it follows that
	$\fma(F'_{[3,1)}) = m^2n$ and
	$\fma(F''_{[3,1)}) = 2m^3n +m^2n^2$
	and hence
	$\fma(F'') =\fma(F''_{[3,0)})= 3m^3n +3m^2n^2+m^2n.$
\end{itemize}
The cost of bracketing from the right grows as $n^3$ and $m^2.$
The opposite holds for the cost of bracketing from the left growing as
$m^3$ and $n^2$.
Linear growth of the discrepancy suggests significant potential 
for further analysis of {\sc Hessian Chain Bracketing}. For example, 
$n=2$ and $m=1$ yield costs of $48\fma$ and $20\fma$
when bracketing from right and left. Further results
presented in Section~\ref{sec:num} suggest that the theoretical savings
also yield corresponding speedups when evaluating the Hessian chains 
numerically. \\
\\
The efficient evaluation of Hessians has been investigated actively 
in the context of AD
since the 1970s \cite{Powell1979OtE}. Particular focus has been set on
the detection \cite{Bhowmick2008APT} and exploitation of structure 
\cite{Goldfarb1984OEo,Hovland1997EDC} and sparsity \cite{Walther2008CSH}.
More recent contributions
include \cite{Gower2012Anf} and \cite{Petra2018OeH}.
To the best of our knowledge, the novelty of this paper's approach to
efficient Hessian accumulation is not violated.

The upcoming material is organized as follows: 
A dynamic programming algorithm for {\sc Hessian Chain Bracketing} is proposed in Section~\ref{sec:dp} including a detailed illustration of the individual steps performed by the algorithm for the simple example introduced above.
Numerical results presented in Section~\ref{sec:num} show potential 
reductions of the operations count over the obvious approaches (bracketing 
from left or right) by factors of ten and more on a set of sample problems
of growing size. The savings are shown to translate into actual improvements
in runtime. All results can be reproduced with the open-source reference
implementation presented in the appendix. Conclusions drawn 
in Section~\ref{sec:concl} are complemented with remarks on ongoing and future
research and development.
Supporting material is collected in the appendix.
{\sc Hessian Accumulation} is shown to be NP-complete in Section~\ref{a1}.
A sample session of our proof-of-concept implementation of the dynamic
programming algorithm from Section~\ref{sec:dp} can be found in Section~\ref{a2}.

\section{Dynamic Programming} \label{sec:dp}

The number of bracketings of $F=F_{[q,0)}$ is known to be 
equal to $\frac{1}{q} \binom{2(q-1)}{q-1}=\frac{(2q-2)!}{q!(q-1)!}$ 
\cite{Catalan}, which grows 
exponentially with $q.$ Subproblems are defined by recursive bisection
as
$$
F_{[i, k)}=F_{[i, j)} \circ F_{[j, k)} 
=(F_i \circ \ldots \circ F_{j+1}) \circ (F_j \circ \ldots \circ F_k) 
$$
for $i=1,\ldots,q,$ $i-k=1,\ldots,q$ and $k<j<i.$
An \fma-optimal bracketing of the Jacobian chain product in 
Equation~(\ref{e1.2}) can be computed by dynamic programming. 
Solutions to subproblems of growing length $i-k$ are tabulated as
$$
\fma(F'_{[i,k)}) =  \underset{k < j < i}{\min} ( 
\fma(F'_{[i,j)})
+  
\fma(F'_{[j,k)}) + 
\fma(F'_{[i,j)}\cdot F'_{[j,k)}) 
) \; .
$$
The tabulated costs are used for the minimization of the numbers of $\fma$ 
operations required for the computations of the Hessians
$F''_{[i,k)} \in \R^{n_i \times n_k \times n_k}$ as follows:
\begin{align*}
	\fma(F''_{[i,k)}) =  \underset{k  < j <i}{\min} &( 
\fma(F'_{[i,j)})+\fma(F''_{[j,k)})+\fma(F'_{[i,j)} \cdot F''_{[j,k)}) \\
	&+ \fma(F''_{[i,j)})+\fma(F'_{[j,k)})+\fma(F''_{[i,j)} \cdot F'_{[j,k)} \otimes F_{[j,k)}
) \; .
\end{align*}

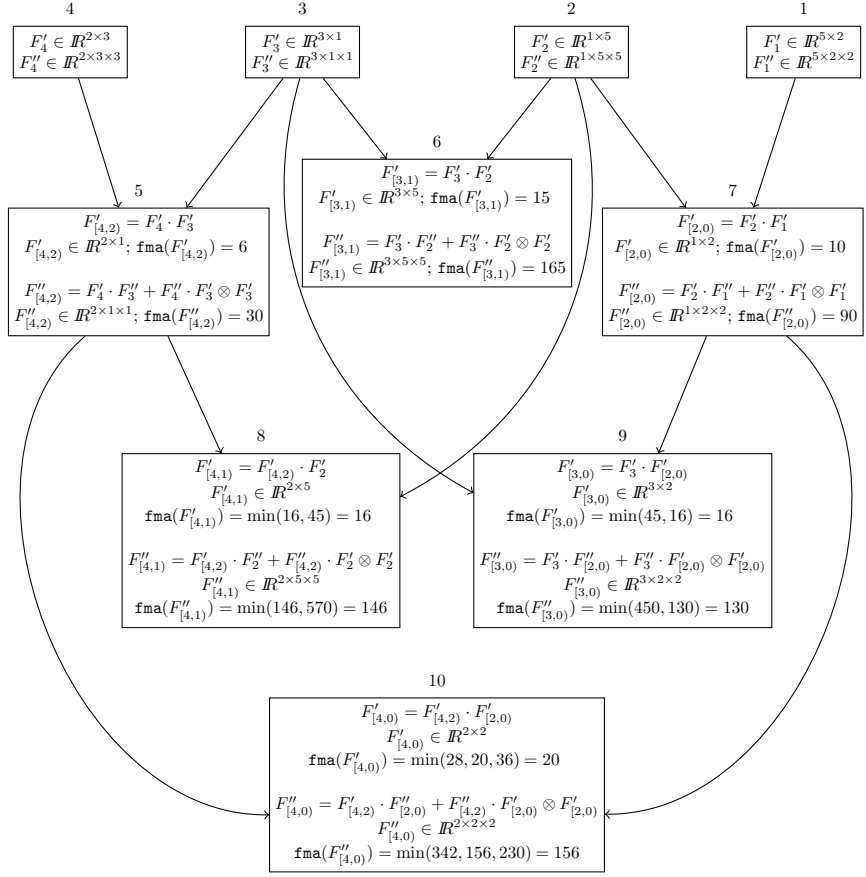
\begin{figure*}
\begin{minipage}[c]{\linewidth}
\begin{tikzpicture}[->,scale=.65, transform shape, rectangle]

\tikzstyle{every node}=[draw,rectangle,align=center,minimum width=1cm, minimum height=0.7cm]
          \node [label={\normalsize{\text{4}}}] (1) at (0,14) {
	  $F'_4\in \R^{2\times 3}$ \\
	  $F''_4\in \R^{2\times 3\times 3}$ 
	  };
	  \node [label={\normalsize{\text{3}}}] (2) at (4.75,14) {
		  $F'_3\in \R^{3\times 1}$ \\
		  $F''_3\in \R^{3\times 1\times 1}$ 
		  };
          \node [label={\normalsize{\text{2}}}] (3) at (10.25,14) {
		  $F'_2\in \R^{1\times 5}$ \\
		  $F''_2\in \R^{1\times 5\times 5}$ 
		  };
          \node [label={\normalsize{\text{1}}}] (4) at (15,14) {
		  $F'_1\in \R^{5\times 2}$ \\
		  $F''_1\in \R^{5\times 2\times 2}$ 
		  };
          \node [label={\normalsize{\text{5}}}] (5) at (1.4,9.5) {
	  $F'_{[4,2)} = F'_4\cdot F'_3$ \\
		  $F'_{[4,2)} \in \R^{2\times 1};$ $\fma(F'_{[4,2)}) = 6$ \vspace{.3cm} \\
	  $F''_{[4,2)} = F'_4\cdot F''_3 + F''_4\cdot F'_3\otimes F'_3$ \\
		  $F''_{[4,2)}\in \R^{2\times 1\times 1};$ $\fma(F''_{[4,2)})=30$ 
	  };
          \node [label={\normalsize{\text{6}}}] (6) at (7.5,10.5) {
		  $F'_{[3,1)} = F'_3\cdot F'_2$ \\
		  $F'_{[3,1)}\in \R^{3\times 5};$ $\fma(F'_{[3,1)}) = 15$ \vspace{.3cm} \\
		  $F''_{[3,1)} = F'_3\cdot F''_2 + F''_3\cdot F'_2\otimes F'_2$\\
		  $F''_{[3,1)}\in \R^{3\times 5\times 5};$ $\fma(F''_{[3,1)})=165$  
		  };
          \node [label={\normalsize{\text{7}}}] (7) at (13.55,9.5) {
		  $F'_{[2,0)} = F'_2\cdot F'_1$ \\
		  $F'_{[2,0)}\in \R^{1\times 2};$ $\fma(F'_{[2,0)}) = 10$ \vspace{.3cm} \\ 
		  $F''_{[2,0)} = F'_2\cdot F''_1 + F''_2\cdot F'_1\otimes F'_1$ \\
		  $F''_{[2,0)}\in \R^{1\times 2\times 2};$  
		  $\fma(F''_{[2,0)})=90$ 
		  };
          \node [label={\normalsize{\text{8}}}] (8) at (3.9,4) {
		  $F'_{[4,1)} = F'_{[4,2)}\cdot F'_2 $ \\
		  $F'_{[4,1)}\in \R^{2\times 5}$ \\ 
		  $\fma(F'_{[4,1)}) = \min(16,45)=16$  \vspace{.3cm} \\ 
		  $F''_{[4,1)} = F'_{[4,2)}\cdot F''_2 + F''_{[4,2)}\cdot F'_2\otimes F'_2$ \\
		  $F''_{[4,1)}\in \R^{2\times 5\times 5}$ \\ 
		  $\fma(F''_{[4,1)})=\min(146,570)=146$  
}; 
          \node [label={\normalsize{\text{9}}}] (9) at (11.3,4) {
		  $F'_{[3,0)} =    F'_3\cdot F'_{[2,0)}$ \\
		  $F'_{[3,0)}\in \R^{3\times 2}$ \\ 
		  $\fma(F'_{[3,0)}) = \min(45,16)=16$ \vspace{.3cm} \\ 
		  $F''_{[3,0)} = F'_3\cdot F''_{[2,0)} + F''_3\cdot F'_{[2,0)}\otimes F'_{[2,0)}$ \\
		  $F''_{[3,0)}\in \R^{3\times 2\times 2}$ \\ 
		  $\fma(F''_{[3,0)})=\min(450,130)=130$ 
		  };
          \node [label={\normalsize{\text{10}}}] (10) at (7.5,-1) {
		  $F'_{[4,0)} = F'_{[4,2)}\cdot F'_{[2,0)}$ \\
		  $F'_{[4,0)}\in \R^{2\times 2}$ \\ 
		  $\fma(F'_{[4,0)}) = \min(28,20,36)=20$ \vspace{.3cm} \\ 
		  $F''_{[4,0)} =    F'_{[4,2)}\cdot F''_{[2,0)} + F''_{[4,2)}\cdot F'_{[2,0)}\otimes F'_{[2,0)}$ \\
		  $F''_{[4,0)}\in \R^{2\times 2\times 2}$ \\ 
		  $\fma(F''_{[4,0)})=\min(342,156,230)=156$ 
		  };
    \begin{scope}[on background layer]
    \path[every node/.style={font=\sffamily}]
        (1) edge node [right, scale=0.75] {} (5)
    	(2) edge node [left, scale=0.75] {} (5)
    	(2) edge node [right, scale=0.75] {} (6)
        (3) edge node [left, scale=0.75] {} (6)
    	(3) edge node [right, scale=0.75] {} (7)
    	(4) edge node [left, scale=0.75] {} (7)
        (5) edge node [right, scale=0.75] {} (8)
    	(3) edge [bend left=40] node [left, scale=0.75,pos=0.2] {} (8)
   	(2) edge [bend right=39.5] node [right, scale=0.75,pos=0.2] {} (9)
    	(7) edge node [left, scale=0.75] {} (9)
    	(5) edge [bend right=70] node [right, scale=0.75] {} (10)
    	(7) edge [bend left=70] node [left, scale=0.75] {} (10) ;
      \end{scope}
\end{tikzpicture}
\end{minipage}
\vspace{.5cm}
\caption{Dynamic programming for {\sc Hessian Chain Bracketing}:
The algorithm is visualized as a directed acyclic graph 
	for $F=F_4 \circ F_3 \circ F_2 \circ F_1$ with 
	$F_4 : \R^3 \rightarrow \R^2,$
	$F_3 : \R \rightarrow \R^3,$
	$F_2 : \R^5 \rightarrow \R,$
	$F_1 : \R^2 \rightarrow \R^5.$
	Vertices 1 to 4 (5 to 7 $|$ 8 to 9) correspond to subchains of length 
	one (two $|$ three). The optimal bracketing is represented by vertex 10.
Vertices contain information on the computation of Jacobians and Hessians of the
respective subchains corresponding to the optimal bracketing. 
The dimensions of the resulting Jacobians and Hessian are stated as well as
the numbers of \fma\ required for their computation. Edges visualize split 
positions by linking a chain to its two subchains according to an optimal bracketing. For example,
	$F''_{[4,0)}$ is computed optimally based on the bracketing 
	$(F_4 \circ F_3) \circ (F_2 \circ F_1)$ at the cost of
	$\fma(F''_{[4,2)}) + \fma(F'_{[4,2)})+\fma(F''_{[2,0)}) + \fma(F'_{[2,0)})+n_4 n_2 n_0^2+n_4 n^2_2 n_0 + n_4 n_2 n_0^2=30+6+90+10+8+4+8=156 \fma.$
}
\label{fig:dyn}
\end{figure*}

Correctness of the algorithm follows immediately from the {\em optimal 
substructure} and {\em overlapping subproblems} properties 
\cite{Bellman1957DP} exhibited by both {\sc Jacobian} and 
{\sc Hessian Chain Bracketing}.
\paragraph*{Example} We use the same example 
as in Section~\ref{sec:1} with $n=2$ and $m=1$
for illustration of the individual steps of the dynamic programming algorithm,
that is,
$F=F_3 \circ F_2 \circ F_1$
such that $F_1,F_3 : \R^2 \rightarrow \R$
and $F_2 : \R \rightarrow \R^2.$ Again, and without loss of generality we
assume elemental Jacobians and Hessians to be dense. 
The number of \fma\ required for the product of matrix with a vector is
invariant with respect to potential symmetry of the matrix. Hence,
the exploitation of likely
symmetry of the Hessians does not lead to a reduction in the \fma-cost.

For a function composition of length three there are only two choices 
corresponding to bracketing from the left at the computational cost of
20\fma\ or bracketing from the right at 48\fma. The algorithm favors the
former as the result of performing the following steps:

The optimal bracketings of all Jacobian subchains are computed as
\begin{align*}
	\fma(F'_{[2,0)})&=n \cdot m \cdot n = m \cdot n^2 =4; \;
\fma(F'_{[3,1)})=m \cdot n \cdot m = m^2 \cdot n =2\\
	\fma(F'_{[3,0)})&=\min( 
\fma(F'_{[2,0)})+\fma(F'_3 \cdot F'_{[2,0)}), \fma(F'_{[3,1)}) +\fma(F'_{[3,1)} \cdot F'_1) 
) \\
&=\min(m \cdot n^2 + m \cdot n \cdot n, m^2 \cdot n + m \cdot m \cdot n) \\
&=\min(2 \cdot m \cdot n^2, 2 \cdot m^2 \cdot n ) 
=\min(8, 4) =4 \; . 
\end{align*}
The whole chain for
is evaluated with minimal $\fma$ cost of four as $F'=(F'_3 \cdot F'_2) \cdot F'_1.$ 

Dynamic programming for {\sc Hessian Chain Bracketing} yields costs for the two 
subchains of length two as
\begin{align*}
\fma&(F''_{[2,0)}) =  
\fma(F'_2 \cdot F''_1)+ \fma(F''_2 \cdot F'_1 \otimes F'_1) \\
&=n \cdot m \cdot n \cdot n + n \cdot m \cdot n \cdot (m+n) \\
	&=m \cdot n^3 + m \cdot n^2 \cdot (m+n) = 8+12=20\\
\fma&(F''_{[3,1)}) =  
\fma(F'_3 \cdot F''_2)+ \fma(F''_3 \cdot F'_2 \otimes F'_2) \\
&=m \cdot n \cdot m \cdot m + m \cdot n \cdot m \cdot (m+n) \\
&=m^3 \cdot n + m^2 \cdot n \cdot (m+n) = 2+6=8\\
\intertext{which are looked up during the optimization of $\fma(F''_{[3,0)})$ as}
	\fma(F''_{[3,0)}) &=  \underset{0 < j <3}{\min} ( 
\fma(F'_{[3,j)} \cdot F''_{[j,0)})+ \fma(F''_{[3,j)} \cdot F'_{[j,0)}\otimes F'_{[j,0)}) 
) \displaybreak[0]\\
={\min} ( \\
&\fma(F'_{[3, 1)} \cdot F''_1)+ \fma(F''_{[3,1)} \cdot F'_1 \otimes F'_1),\\
&\fma(F'_3 \cdot F''_{[2,0)})+ \fma(F''_3 \cdot F'_{[2,0)} \otimes F'_{[2,0)}) \\
) \displaybreak[0]\\
={\min} ( \\
&\fma(F'_{[3,1)}) + \fma(F''_1) + m \cdot m \cdot n^2 \\
&+ \fma(F''_{[3,1)}) +\fma(F'_1) + m \cdot m \cdot n \cdot (m+n),\\
&\fma(F'_3) + \fma(F''_{[2,0)}) + m \cdot n \cdot n^2\\
&+ \fma(F''_3) + \fma(F'_{[2,0)})+m \cdot n \cdot n \cdot (n + n) \\
) \displaybreak[0]\\
={\min} ( \\
&(m^2 \cdot n) + 0 + (m^2 \cdot n^2) \\
&+ (m^3 \cdot n + m^2 \cdot n \cdot (m+n)) \\
&+0+ (m^2 \cdot n \cdot (m+n)),\\
&0 + (m \cdot n^3 + m \cdot n^2 \cdot (m+n)) + m \cdot n^3\\
&+ 0 + n^2+2 \cdot m \cdot n^3 \\
) \\
	=\min&(20,48)=20 \; .
\end{align*} 
This result validates the observations made in Section~\ref{sec:1}.

\section{Case Studies} \label{sec:cs}
A detailed illustration of the dynamic programming algorithm 
for the composite function
$
F=F_4 \circ F_3 \circ F_2 \circ F_1
$
can be found in Figure \ref{fig:dyn} with further comments provided in the 
corresponding caption. The solution to {\sc Hessian Chain Bracketing} for
		$F_1 : \R^2 \rightarrow \R^5,$
		$F_2 : \R^5 \rightarrow \R,$
		$F_3 : \R \rightarrow \R^3,$
		$F_4 : \R^3 \rightarrow \R^2$ is computed based on
		$$
F=(F_4 \circ F_3) \circ (F_2 \circ F_1)
		$$
		with a total cost of $156\fma$ required for the accumulation 
		of the Hessian $F''.$
		Again and without loss of generality, all elemental Jacobians and Hessians are regarded as dense.

As a real-world case study we consider 
the LIBOR\footnote{London Interbank Offered Rate} market model
introduced in \cite{Brace1997TMM}
and used in \cite{Giles2006Saf}
as illustration of the benefits of adjoint AD for simulations in finance.
Over recent years adjoint AD has gained significant importance in computational
finance driven mainly by increasing gradient sizes in the context of XVA
calculations and documented by a large number of related publications, e.g,
\cite{Pfadler2015CSo,Lu2016TXo}. Considerable effort has been going into the 
training of surrogate models based on artificial neural networks (ANN)
\cite{Huge2020Dml}.

The LIBOR sample code simulates the evolution of the LIBOR rates for a 
portfolio of
swaptions with given swap rates and maturities. 
As in \cite{Giles2006Saf}, 
swaps of the floating forward
rate $L \in \R^n$ and a given fixed swap rate are considered for $n=80$. 
Monte Carlo simulation with a normally distributed random variable
$Z \in \R^{p \times m}$ performs $p$ path
calculations evolving $L=L(t)$ for $m$ time steps to the target time $t=T$ 
and starting from a given initial state $L(0)$. 
Refer to \cite{Glasserman2003MCM} for further discussion of the mathematical
details behind the LIBOR market model. All numerical results obtained by
our implementation were validated against the implementation used in
\cite{Giles2006Saf} and available from Giles' website\footnote{\tt people.maths.ox.ac.uk/gilesm/codes/libor\_AD} at the University of Oxford, UK.

On the given computer the run time of $p=10^4$ primal Monte Carlo path 
simulations is $1.9s$. We consider the accumulation of the Hessian
$\frac{d^2L(T)}{dL(0)^2} \in \R^{80 \times 80 \times 80}$ based on a surrogate 
model in form of an ANN with 11 layers and 80 nodes per layer
trained to $99\%$ accuracy in terms of mean squared error. Subsequent pruning
eliminates insignificant nodes from hidden layers
as described in \cite{Afghan2020IAS} and
based on the results an 
interval adjoint significance analysis introduced in \cite{Vassiliadis2016TSA}.
A layered function is generated as in Equation~(\ref{e1.1}) with $q=11$ and
$n_0=80,$
$n_1=32,$
$n_2=65,$
$n_3=64,$
$n_4=55,$
$n_5=46,$
$n_6=n_7=49,$
$n_8=53,$
$n_9=62,$
$n_{10}=48,$
$n_{11}=80.$ The pruned ANN preserves the 
$99\%$ target accuracy on the given test set.

Based on the measured primal runtime of $1.9s$ the accumulation of the Hessian
in second-order tangent mode of AD is estimated to take approximately 
$1.5 \cdot 80^2 \cdot 1.9=18,240s$ or 5 hours. The factor $1.5$ is due to the 
overhead of a tangent (directional derivative) propagation induced by our
AD library dco/c++ \cite{AIB-2016-08}. A total of $80^2$ tangents need to be 
evaluated. 

The runtime of the surrogate is negligible (a few milliseconds; $ms$). 
So is the cost of evaluation of the elemental Hessians (a few seconds). Our 
runtime measurements assume the latter to be given. Different bracketing of 
Equation~(\ref{e1.1}) are compared. Bracketing from the left [right] performs
$388,844,400\fma$ [$517,283,120\fma$] in $855ms$ [$1,125ms$]. A greedy 
heuristic based on locally optimal decisions results in $298,631,368\fma$ 
taking $638ms.$ Dynamic programming yields an optimal bracketing with 
$149,061,728\fma$ performed in $311ms.$ The reduction in the number of $\fma$
by a factor of almost three carries over to the runtime. The optimal bracketing
evaluates the Hessian based on 
$$
F=(F_{11} \circ (F_{10} \circ (F_{9} \circ (F_{8} \circ (F_{7} \circ (F_{6} \circ (F_{5} \circ (F_{4} \circ (F_{3} \circ F_{2}))))))))) \circ F_{1} \; .
$$
All results can be reproduced (runtimes qualitatively) using the reference
implementation described in the appendix.

\section{Further Numerical Results} \label{sec:num}

\begin{table*}[t]
	\caption{Random Test Cases}
 \begin{tabular}{|c c c c c|} 
 \hline
	 $q$ & optimized bracketing & bracketing from left & ... from right & $\fma_\text{rel}$ \\ [0.5ex] 
 \hline
 3 & \num[group-separator={,}]{20} & \num[group-separator={,}]{20} & \num[group-separator={,}]{48} & \bf 1 \\ 
 4 & \num[group-separator={,}]{156} & \num[group-separator={,}]{342} & \num[group-separator={,}]{230}  & \bf 1.47 \\ 
 5 & \num[group-separator={,}]{1218} & \num[group-separator={,}]{2210} & \num[group-separator={,}]{1860} &\bf 1.52 \\ 
 10 & \num[group-separator={,}]{11220} & \num[group-separator={,}]{53118} & \num[group-separator={,}]{20952} & \bf 1.86 \\ 
 15 & \num[group-separator={,}]{10200} & \num[group-separator={,}]{217555} & \num[group-separator={,}]{51152} & \bf 5.01 \\
 20 & \num[group-separator={,}]{56830} & \num[group-separator={,}]{1057665} & \num[group-separator={,}]{1053117} & \bf 18.53 \\
 25 & \num[group-separator={,}]{286366} & \num[group-separator={,}]{3953376} & \num[group-separator={,}]{1345312} & \bf 4.69 \\
 30 & \num[group-separator={,}]{195620} & \num[group-separator={,}]{1655596} & \num[group-separator={,}]{8615838} & \bf 8.46 \\ 
 35 & \num[group-separator={,}]{614499} & \num[group-separator={,}]{23461452} & \num[group-separator={,}]{4440564} & \bf 7.22 \\ 
 40 & \num[group-separator={,}]{2254794} & \num[group-separator={,}]{24388365} & \num[group-separator={,}]{66718064} & \bf 10.81 \\
 45 & \num[group-separator={,}]{1787606} & \num[group-separator={,}]{28170189} & \num[group-separator={,}]{139760800} & \bf 15.75 \\
 50 & \num[group-separator={,}]{8271082} & \num[group-separator={,}]{170383616} & \num[group-separator={,}]{65760913} & \bf 7.95 \\ [1ex] 
 \hline
 \end{tabular}
\label{table:1}
\end{table*}

\begin{table*}[t]
	\caption{Runtimes for Larger Random Test Cases}
\begin{tabular}{|c c c c c|} 
 \hline
	 $q$ & optimized bracketing & best unidirectional bracketing & $\fma_\text{rel}$ & $t_\text{rel}$\\ [0.5ex] 
 \hline
60 & \num[group-separator={,}]{37989141} & \num[group-separator={,}]{331166304} & 8.71 & \bf 5.54\\ 
70 & \num[group-separator={,}]{9241074} & \num[group-separator={,}]{554021568}  & 59.95 & \bf 14.05\\ 
80 & \num[group-separator={,}]{33796544} & \num[group-separator={,}]{200622919} & 5.93 & \bf 2.44\\
90 & \num[group-separator={,}]{19058174} & \num[group-separator={,}]{251632865} & 13.20 &\bf  2.36\\
100 & \num[group-separator={,}]{22951156} & \num[group-separator={,}]{2313419043} & 100.79 & \bf 16.20\\
150 & \num[group-separator={,}]{502245226} & \num[group-separator={,}]{16928546112} & 33.70 & \bf 13.53\\
200 & \num[group-separator={,}]{816938109} & \num[group-separator={,}]{4620627490} & 5.65 & \bf 1.42\\[1ex] 
 \hline
 \end{tabular}
\label{table:2}
\end{table*}

	Table~\ref{table:1} lists the results obtained by applying 
	the dynamic programming heuristic for {\sc Hessian Chain 
	Bracketing} to chains of elemental functions of growing length $q.$ 
	The latter also serves as an upper bound for the randomly generated 
	dimensions of domains and images of the individual elemental functions.
	We compare the numbers of \fma\ required for the accumulation of the
	Hessian when bracketing from the left or from the right with
	the numbers resulting from optimized bracketing. The factor
	quantifying the improvement due to optimized bracketing 
	over the better out of the uniform bracketings is shown in the last 
	column. Relative savings in the \fma\ count of up to eighteen can
	be observed.

	Savings in the number of \fma\ required for the accumulation of the 
	Hessian can be expected to yield adequate reductions in runtime.
	A set of larger problem instances is presented for this 
	purpose in Table~\ref{table:2}. 
	Relative savings in the \fma\ count of up to one hundred result in
	speedups of up to sixteen as shown in the last column. Our reference implementation is not tuned 
	for speed. It uses Eigen\footnote{\tt https:://eigen.tuxfamily.org} 
 for the matrix products. While we consider this approach to be a realistic 
 scenario further optimization is likely to yield even better efficiency. 
For example, the use of GPGPU has been shown to be beneficial
\cite{Gremse2015GAS}.

\section{Conclusion and Outlook} \label{sec:concl}

The results presented in this paper are promising. Reductions in the 
number of \fma\ required for the accumulation of Hessian tensors yield
corresponding speedups. Nevertheless, 
significant effort is required to bridge the present gap to seamless 
integration into software tools for AD. A matrix-free formulation in particular
is necessary to handle computationally complex elemental functions similar
to the first-order scenario investigated in \cite{Naumann2020OoG}. The 
assumption about elemental Hessians being given turns out to be infeasible
in many practical applications. ANN represent an exception as 
differentiation of the individual layers often turns out to be relatively 
straightforward.

Dynamic programming for {\sc Jacobian} and {\sc Hessian Chain Bracketing} 
generalizes to arbitrary order. So does the proof of NP-completeness of 
{\sc Jacobian} and {\sc Hessian Accumulation} as shown in 
\cite{Naumann2021OtC}. The obvious discrepancies between the respective 
formulations give rise to further ongoing investigations into the 
combinatorics induced by the chain rule of differentiation.

\appendix

\section{Complexity Analysis} \label{a1}

The proof of Theorem~\ref{np} builds on the same fundamental ideas as similar
arguments presented in \cite{Naumann2008OJa}.
It uses reduction from {\sc Ensemble Computation} which was shown to be
NP-complete in \cite{Garey1979CaI}.

Consider an arbitrary instance $(A,C,K)$ of {\sc Ensemble Computation} and a bijection
$A \leftrightarrow \tilde{A},$ where $\tilde{A}$ consists of $|A|$ mutually
distinct primes.
A corresponding bijection
$C \leftrightarrow \tilde{C}$ is implied.
Create an extension $(\tilde{A} \cup \tilde{B},\tilde{C},K+|\tilde{B}|)$
by adding unique entries from a sufficiently large set $\tilde{B}$
of primes not in $\tilde{A}$ to the $\tilde{C}_j$ such that they all have
the same cardinality $q$. Note that a solution for this extended
instance of {\sc Ensemble Computation} implies a solution of the original instance of {\sc Ensemble Computation} as each
entry of $\tilde{B}$ appears exactly once.

Fix the order of the elements of the
$\tilde{C}_j$ arbitrarily yielding
$\tilde{C}_j=(\tilde{c}^j_i)_{i=1}^q$ for $j=1,\ldots,|\tilde{C}|.$
Let
$$
F : \R \rightarrow \R^{|\tilde{C}|} : \Y=\Z_q=F(x)
$$
with $F=F_q \circ F_{q-1} \circ \ldots \circ F_1$
defined as
\begin{align*}
F_1 &: \R \rightarrow \R^{|\tilde{C}|} : \; \Z_1=F_1(x) : \;
z^1_j=\frac{\tilde{c}^j_1}{2} \cdot x^2 \\
F_i &: \R^{|\tilde{C}|} \rightarrow \R^{|\tilde{C}|} : \; \Z_i=F_i(\Z_{i-1}) :   \;
z^i_j=\tilde{c}^j_i \cdot z^{i-1}_j
\end{align*}
yielding
\begin{align*}
	F'_1&=\left (\tilde{c}^j_1 \cdot x\right) \in \R^{|\tilde{C}|}=\R^{{|\tilde{C}|} \times 1} \;\; \text{and} \;\;
F''_1=\left (\tilde{c}^j_1 \right) \in \R^{|\tilde{C}|}=\R^{{|\tilde{C}|} \times 1 \times 1}
\end{align*}
as well as diagonal
Jacobians $$F'_i=(d^i_{j,k}) \in \R^{{|\tilde{C}| \times |\tilde{C}|}} \; ,$$
where
$$
d^i_{j,k}=
\begin{cases}
\tilde{c}^j_i & \text{if}~j=k \\
0 & \text{otherwise,} \\
\end{cases}
$$
and vanishing Hessians $F''_i=0$
for $j=1,\ldots,|\tilde{C}|$ and $i=2,\ldots,q.$
Equation~(\ref{e1.3}) simplifies to
$$
F^{\prime \prime}=\prod_{i=2}^{q} F_{i}^{\prime} \cdot F_{1}^{\prime \prime} \; .
$$
According to the fundamental theorem of arithmetic \cite{Gauss1801DA}
the elements of $\tilde{C}$
correspond to unique (up to commutativity of scalar multiplication)
factorizations of the $|\tilde{C}|$ nonzero entries of $F'' \in \R^{|\tilde{C}|}=\R^{{|\tilde{C}|} \times 1 \times 1}.$
This uniqueness property extends to arbitrary subsets
of the $\tilde{C}_j$ considered during the exploration of the search space
of the {\sc Hessian Accumulation} problem.
A solution implies a solution of the associated extended
instance of {\sc Ensemble Computation} and, hence, of the original instance
of {\sc Ensemble Computation}.

A proposed solution for {\sc Hessian Accumulation} is easily
validated by counting the at
most $|\tilde{C}|\cdot q$ scalar multiplications performed.

\section{Implementation} \label{a2}

An open-source reference implementation is provided
for easy reproduction of our computational results; see 
\begin{center}
	\tt git@github.com:un110076/HessianChainBracketing.git \; .
\end{center}
The software consists of three separate executables resulting from
implementations given as three C++ source files. 
Problem instances are generated randomly 
by \texttt{generate.exe} for given length of the chain and upper
bound on the dimensions of domains and images of the elemental functions.
The resulting text file serves as input for \texttt{solve.exe} which computes
a solution for the corresponding instance of (dense) {\sc Hessian Chain 
Bracketing}. Both the problem formulation and the solution can be passed to 
\texttt{run.exe} to perform the numerical evaluation of the
Hessian chain product for given randomly initialized elemental Jacobians and 
Hessians. Eigen is expected to be installed in \texttt{./Eigen}.
The code has been tested with the 
GNU C++ compiler under Linux. A \texttt{Makefile} is provided. Essential
information on how to build and run the software is given in
\texttt{README.md}.

A sample session could proceed as follows:
\begin{enumerate}
	\item Running 
		\begin{verbatim}
		generate.exe 4 4 
		\end{verbatim}
		might yield the output
		\begin{lstlisting}
4
5 2
1 5
3 1
2 3
		\end{lstlisting}
corresponding to the example from Figure~\ref{fig:dyn}.
The chain $F_4 \circ F_3 \circ F_2 \circ F_1$ of length four (first line) 
consists of elemental functions 
		$F_1 : \R^2 \rightarrow \R^5$ (line two),
		$F_2 : \R^5 \rightarrow \R$ (line three),
		$F_3 : \R \rightarrow \R^3$ (line four),
		$F_4 : \R^3 \rightarrow \R^2$ (line five).
Let this output be stored in \texttt{problem.txt}.
\item The dynamic programming algorithm is executed as illustrated in 
Figure~\ref{fig:dyn} by running
		\begin{verbatim}
		solve.exe problem.txt 
		\end{verbatim}
Diagnostic output is generated.
		\begin{lstlisting}[basicstyle=\small]
left bracketing fma = 342   
right bracketing fma = 230   
heuristic bracketing fma = 156
optimized bracketing fma = 156 

Dynamic Programming Table:
fma(F''(1,0))=90;  split before 1; dim(F''(1,0))=1x2x2
fma(F''(2,1))=165; split before 2; dim(F''(2,1))=3x5x5
fma(F''(2,0))=130; split before 2; dim(F''(2,0))=3x2x2
fma(F''(3,2))=30;  split before 3; dim(F''(3,2))=2x1x1
fma(F''(3,1))=146; split before 2; dim(F''(3,1))=2x5x5
fma(F''(3,0))=156; split before 2; dim(F''(3,0))=2x2x2
	\end{lstlisting}
The number
of \fma\ required by the optimized bracketing is compared with the numbers
resulting from uniform bracketing from the left and from the right as well as
with the result of the greedy heuristic.
Moreover, the optimized bracketing is stored in a text file
\texttt{solution.txt} as follows:
		\begin{lstlisting}
3 3 2
1 1 0
3 2 0
		\end{lstlisting}
Visiting the lines in reverse order we find that
the first split position is set before $F_3$ yielding
$(F_4 \circ F_3) \circ (F_2 \circ F_1).$ The remaining
two lines indicate (unique) split positions before $F_2$ and 
before $F_4$ within the two subchains (of length two). 
\item Passing both {\tt problem.txt} and {\tt solution.txt} as command
line arguments to \texttt{run.exe} as
		\begin{verbatim}
		run.exe problem.txt solution.txt heuristic_solution.txt
		\end{verbatim}
		run times for the numerical
evaluation of the uniform bracketings are compared with the
run time of computing the Hessian based on the optimized bracketing yielding, 
for example,
		\begin{lstlisting}[basicstyle=\small]
Elapsed time (in microseconds): 
left bracketing: 69
right bracketing: 52
heuristic bracketing: 48
optimized bracketing: 48
		\end{lstlisting}
Obviously, the numbers become more reliable for larger problems. 
\end{enumerate}

\end{document}